\documentclass[paper,11pt]{article}
\pdfoutput=1

\usepackage[pdftex,bookmarks,colorlinks]{hyperref}
\usepackage[pdftex]{graphicx}
\usepackage[amssymb]{SIunits}
\usepackage{mathrsfs,mathcomp}
\usepackage{fancyhdr}
\fancypagestyle{plain}{
\fancyhead[CO]{\footnotesize{Candidate video for the 29th Annual Gallery of Fluid Motion\\ 64rd Annual Meeting of the American Physical Society, Division of Fluid Dynamics \\Baltimore, MD, 20-22 Nov 2011}}
}

\begin{document}

\title{Freezing singularities in water drops}

\author{Oscar R. Enr\'iquez\footnote{oscarenriquez@gmail.com} , \'Alvaro G. Mar\'in, Koen G. Winkels,\\
Jacco H. Snoeijer\\
\emph{Physics of Fluids, University of Twente}\\
}

\maketitle

\begin{abstract}

In this fluid dynamics video we show how a drop of water ($D\approx\unit{2}{\milli\meter}$) freezes into a singular shape when deposited on a cold surface ($T=-20\tccelsius$). The process of solidification can be observed very clearly due to the change in refraction when water turns into ice. The drop remains approximately spherical during most of the process, with a freezing front moving upwards and smoothly following the interface. However, at the final stage of freezing, when the last cap of liquid turns into ice, a singular tip develops spontaneously. Interestingly, the sharp tip of the ice drop acts as a preferential site for deposition of water vapour, and a beautiful ``tree'' of ice crystals develops right at the tip. The tip singularity attracts the vapour in analogy to a sharp lightning rod attracting lightning.

\end{abstract}

\section{Technical Details of the submitted video}

\begin{figure}[h]
\centering
\label{fig1}
\includegraphics[width=0.6\textwidth]{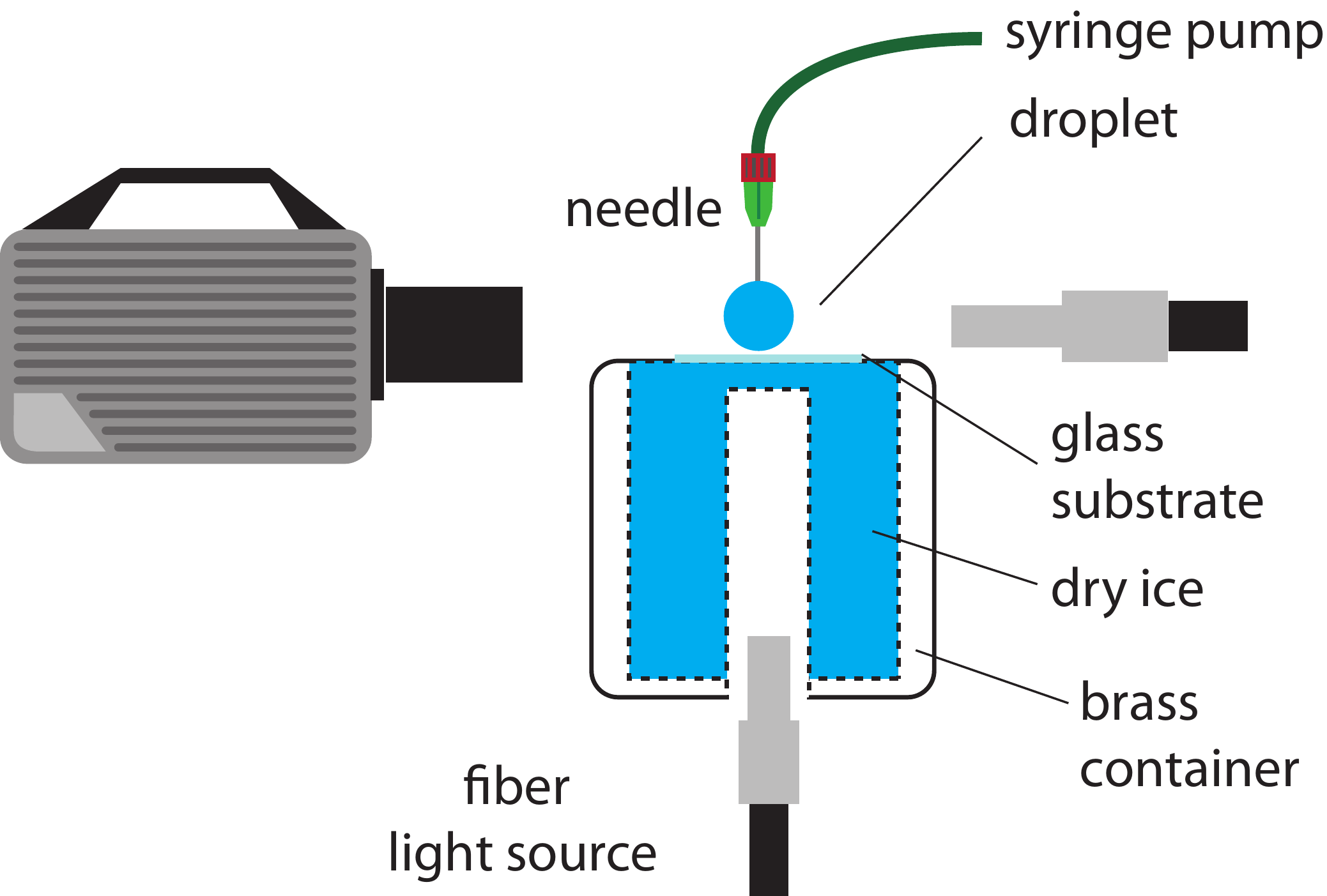}
\caption{Illustration of the experimental setup.}
\end{figure}

The experimental setup (fig. \ref{fig1}) consisted of a brass container filled with solid carbon dioxide (dry ice). A clean glass slide was placed over the brass container, where a drop ($D \approx \unit{2}{\milli\meter}$) of deionized and degassed water was deposited using a syringe pump.  To increase contrast and observe the freezing front, red food dye was added to the water. The process was recorded from the side using a long distance microscope (VZM1000 Edmund Optics) mounted on a color camera, at a frame rate of 50 frames per second. We used both backlight and bottom light illumination provided by optic fiber lamps. The resolution obtained was $2048\times 1152$ pixels with approximately $ 3\ \mu m /pixel$. We measured the plate temperature near the droplet using a standard thermocouple.

Links to videos: \href{http://stilton.tnw.utwente.nl/people/alvaro/freezingdrop1.mp4}{50 MB} and \href{http://stilton.tnw.utwente.nl/people/alvaro/freezingdrop1.mp4}{11 MB}.

\bibliographystyle{plainnat}

\end{document}